\documentclass[twocolumn,showpacs,preprintnumbers,amsmath,amssymb]{revtex4}
\usepackage{epsfig}
\usepackage{graphicx}
\usepackage{dcolumn}
\usepackage{bm}

\def\ube13{UBe$\rm_{13}$}

\def\bi2212{Bi$\rm_2$Sr$\rm_2$CaCu$\rm_2$O$\rm_8$}
\def\ybi2212{Bi$\rm_2$Sr$\rm_2$YCu$\rm_2$O$\rm_8$}
\def\ycabi2212{Bi$\rm_2$Sr$\rm_2$Ca$\rm_{1-x}$Y$\rm_x$Cu$\rm_2$O$\rm_{8+\delta}$}

\def\y65cabi2212{Bi$\rm_2$Sr$\rm_2$Ca$\rm_{0.35}$Y$\rm_{0.65}$Cu$\rm_2$O$\rm_{8+\delta}$}

\def\Co{CeCoIn$_5$}
\def\Rh{CeRhIn$_5$}

\begin{document}

\title{Field-Tuned Quantum Critical Point in CeCoIn$_5$ Near the Superconducting Upper Critical Field}

\author{F.~Ronning,$^1$ C.~Capan,$^1$ A.~Bianchi,$^2$ R.~Movshovich,$^1$ A.~Lacerda,$^3$ M.~F.~Hundley,$^1$ J.~D.~Thompson,$^1$ P.~G.~Pagliuso,$^4$ J.~L.~Sarrao$^1$ }
\affiliation{$^1$Los Alamos National Laboratory, Los Alamos, New Mexico 87545\\
             $^2$Dresden High Magnetic Field Laboratory, Forschungszentrum Rossendorf, 01314 Dresden, Germany\\
             $^3$National High Magnetic Field Laboratory, Los Alamos, New Mexico 87545\\
             $^4$Instituto de Fisica Gleb Wataghin, UNICAMP, 13083-970, Campinas, Brazil}

\date{\today}

\begin{abstract}
We report a systematic study of high magnetic field specific heat
and resistivity in single crystals of \Co~for the field oriented
in the basal plane~(H$\parallel$ab) of this tetragonal heavy
fermion superconductor. We observe a divergent electronic specific
heat as well as an enhanced A coefficient of the T$^{2}$ law in
resistivity at the lowest temperatures, as the field approaches
the upper critical field of the superconducting transition.
Together with the results for field along the tetragonal
axis~(H$\parallel$c), the emergent picture is that of a magnetic
field tuned quantum critical point which exists in the vicinity of
the superconducting  H$_{c2}^{0}$ despite a variation of a factor
of 2.4 in H$_{c2}^{0}$ for different field orientations. This
suggests an underlying physical reason exists for the
superconducting H$_{c2}^{0}$ to coincide with the quantum critical
field. Moreover, we show that the recovery of a Fermi Liquid
ground state with increasing magnetic field is more gradual,
meaning that the fluctuations responsible for the observed quantum
critical phenomena are more robust with respect to magnetic field,
when the magnetic field is applied in-plane. Together with the
close proximity of the quantum critical point and H$_{c2}^{0}$ in
\Co~for both field orientation, the anisotropy in the recovery of
the Fermi liquid state might constitute an important piece of
information in identifying the nature of the fluctuations that
become critical.
\end{abstract}


\maketitle
Quantum critical points mark the change in the ground state of a
strongly correlated electron system, and the associated quantum
fluctuations have tremendous consequences for the properties of
the system at finite temperatures. Attention has focused on the
heavy fermion superconductor \Co~in the context of quantum
criticality since its discovery~\cite{Petrovic:JPCM-01}.
Superconductivity in this material is not only unconventional
(probably {\it d}-wave~\cite{Roman:prl-01,Izawa:prl-01}) and
Pauli-limited (with the possible presence of a
Fulde-Ferrell-Larkin-Ovchinnikov state at low
temperatures)~\cite{Andrea:prl-02,AndreaFFLO:prl-03,Radovan:Nature-03}
but it is also built out of a normal state displaying Non Fermi
Liquid behavior. Indeed, the normal state is characterized by a
resistivity almost linear in temperature for a decade above
T$_{c}$ in zero field~\cite{Petrovic:JPCM-01}, a specific heat
coefficient diverging logarithmically over a large temperature
range with a similar slope at zero and finite magnetic
field~\cite{Petrovic:JPCM-01,Kim:prb-01}, and a power law behavior
in ac-susceptibility~\cite{Petrovic:JPCM-01,Kim:prb-01} and the
nuclear spin-lattice relaxation rate~\cite{Kohori:prb-01}. All of
this suggests the proximity to an antiferromagnetic instability.
It is important to note that the specific heat is analogous to
UBe$_{13}$~\cite{Ott:prl-84}. Since the entropy is conserved
between the zero field superconducting state and the anomalous
normal state at H$_{c2}^{0}$, this implies that the mass
enhancement leading to the heavy fermion ground state is
interrupted by the formation of superconductivity
and presumably the same spin fluctuations are responsible for both phenomena.\\

The phase diagram of \Co~turns out to be rather complex, raising
the possibility of one or more quantum critical points. On the one
hand, under pressure the ground state evolves into a conventional
Fermi Liquid, and the effective mass decreases, as evidenced by
resistivity~\cite{Sidorov:prl-02}, specific
heat~\cite{Sparn:PhysicaB-02}, de Haas-van
Alphen~\cite{Shishido:JPCM-03} and $^{115}$In NQR
measurements~\cite{Kohori:JMMM-04}. Moreover, the similarity in
the pressure dependence of T$_{c}$ in both \Co~and the
isostructural antiferromagnetic compound \Rh~points to the
existence of a pressure-tuned antiferromagnetic quantum critical
point close to ambient pressure in \Co~\cite{Sidorov:prl-02}.\\

On the other hand, systematic transport and thermodynamic
investigations of the normal state at magnetic fields above
H$_{c2}^{\parallel c} \simeq 5$
T~\cite{JP1:prl-03,AndreaQCP:prl-03} have revealed that the ground
state evolves into a Fermi Liquid with increasing field as well,
meaning that pressure is not the only tuning parameter for \Co.
Moreover, for the same reasons as for the pressure phase diagram,
one can speculate that the quantum critical point in the magnetic
field phase diagram is an antiferromagnetic one. Surprisingly, the
critical field is found to be close to the superconducting upper
critical field H$_{c2}^{0}$. Although antiferromagnetic long range
order has \emph{not} been observed in \Co, it has been suggested
that it is avoided due to the formation of the superconducting
ground state, and antiferromagnetic spin fluctuations may still be
responsible for the observed quantum critical behavior at
H$_{c2}^{0}$~\cite{AndreaQCP:prl-03}. Since the superconducting
transition itself is first order at low temperatures, possibly as
a consequence of Pauli
limiting~\cite{Andrea:prl-02,AndreaFFLO:prl-03}, it seemed natural
to exclude the superconducting fluctuations from this picture.
More recently, a report of thermal conductivity above H$_{c2}^{0}$
showed a divergence in the scattering rate exactly identical to
the one obtained from electrical resistivity, ruling out the
possibility of superconducting fluctuations playing an important
role in \Co~\cite{JP2:cond-mat-04}. On the other hand, a recent
study on Sn-doped \Co~shows that the superconducting upper
critical field H$_{c2}^{0}$ is suppressed by Sn-doping exactly in
the same manner as the quantum critical
field~\cite{Eric:unpublished}, suggesting that the presence of a
quantum critical point in the vicinity of H$_{c2}^{0}$ is
\emph{not} a coincidence in \Co. It has also been pointed out in
the pure compound that the quantum fluctuations result in a
sub-linear temperature dependence in resistivity at finite fields,
which is not well understood~\cite{JP2:cond-mat-04}.  To date, the
nature of the critical fluctuations at H$_{c2}^{0}$ is still not
established despite considerable efforts and it adds yet another
mystery to the intimate relationship
between antiferromagnetism and superconductivity in the 115 family.\\

All the above-mentioned work related to quantum critical phenomena
at finite fields has been performed with the magnetic field
applied \emph{parallel} to the tetragonal c-axis, which is the
easy axis of magnetization in the 115 family. Since the upper
critical field is anisotropic, it is important to check if the
phase diagram is similar when the field is applied in the basal
plane, i.e. whether the quantum critical behavior is tied to the
destruction of superconductivity at H$_{c2}$. This is precisely
the motivation of this work. We measured specific heat and
resistivity in single crystals of \Co~for magnetic fields
\emph{perpendicular} to the c-axis, ranging between 12~T and 18~T
and temperatures between 50~mK and 3~K, in the 20~T magnet at the
National High Magnetic Field Laboratory, using a dilution
refrigerator. Specific heat is measured in the same single crystal
for both field orientations, with a quasi-adiabatic heat pulse
technique, so that we can compare this data against the specific
heat data for H$\parallel$c from ref.~\cite{AndreaQCP:prl-03}.
Resistivity is measured in a second single crystal, of good
quality and geometry, with no free In, having a RRR ratio of 111
with a residual resistivity of 0.3~$\mu \Omega$.cm with
H$\parallel$ab. The contacts are made by spotwelding Pt wires with
a geometry such that J$\perp$H and J,H$\parallel$ab, and care was
taken to ensure that there is no self-heating in the sample
created by current at the lowest temperatures. Both specific heat
and resistivity results for the in-plane orientation show a
magnetic field-tuned quantum critical point in the vicinity of the
upper critical field H$_{c2}^{\parallel ab} \simeq$ 11.8~T,
similar to the c-axis results, even though the upper critical
field has increased by a factor of 2.4 as compared to the c-axis.
Moreover, we show that the magnetic field is less effective in
suppressing the critical fluctuations and restoring the Fermi
Liquid behavior in this orientation, as compared to H$\parallel$c.
This is in contrast to the behavior observed in tetragonal
YbRh$_{2}$Si$_{2}$, which is an example of a field tuned
antiferromagnetic quantum critical point. For YbRh$_{2}$Si$_{2}$,
the evolution of the Fermi temperature determined from resistivity
as well as the divergence of the T$^{2}$ term in resistivity as a
function of the reduced field are roughly the same for the two
field orientations~\cite{Gegenwart:prl-02}. Although the
anisotropy factor in the critical field of \Co(2.4) is much
smaller than YbRh$_{2}$Si$_{2}$(11), our results show that the
effective "distance" to the quantum critical point depends on the
orientation of the magnetic field. In light of these observations,
more theoretical work is needed to better understand the nature of
the field-tuned quantum critical point in \Co.

\begin{figure}
\includegraphics[width=3.3in]{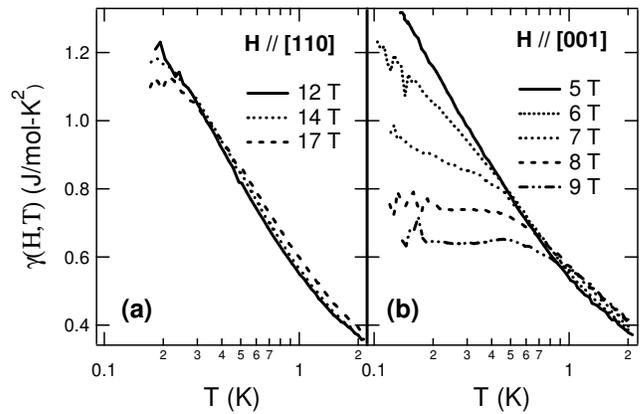}
\caption{\label{Anisotropy-gamma-fig1} The Sommerfeld coefficient
of the electronic specific heat as a function of temperature for
magnetic fields above $H_{c2}^{0}$ oriented in the plane (left
panel) and parallel to c-axis (right panel) on the same crystal.
The evolution from NFL to FL behavior with increasing field is
more gradual when the field is oriented in-plane. The data for
H$\parallel$c is from ref.~\cite{AndreaQCP:prl-03}.}
\end{figure}

Figure~\ref{Anisotropy-gamma-fig1} shows the electronic specific
heat coefficient $\gamma~\equiv$~C$_{el}$/T in the normal state as
a function of temperature, on a semi-logarithmic scale, for the
magnetic field oriented in the plane (left panel) and parallel to
the c-axis (right panel) in the same single crystal. The
electronic contribution is obtained after subtraction of the
nuclear Schottky and lattice contributions from the measured
specific heat~\cite{Roman:prl-01}. In both orientations, the
specific heat is divergent down to the lowest measured temperature
at H$\approx$H$_{c2}$ which is 4.95~T and 11.8~T respectively for
field parallel and perpendicular to the c-axis. Moreover, the two
curves at these fields overlap almost perfectly in the whole
temperature range for the two orientations. However, the evolution
of the specific heat as the magnetic field is increased above
H$_{c2}$ depends on the field orientation. For magnetic fields in
the plane, the specific heat is barely changed over the entire
temperature range when the field is increased up to 17~T,
corresponding to a 44$\%$ relative increase above the critical
field. Both 14~T and 17~T curves show essentially a diverging
specific heat as the temperature decreases, corresponding to Non
Fermi Liquid behavior. Only at 17~T a crossover to a Fermi Liquid
regime, characterized by a constant $\gamma$, can be resolved
around 0.2~K, with a $\gamma$(0.2~K) value reduced to
1.1~J/molK$^{2}$, only 8$\%$ less than its 0.2~K value at
H$_{c2}$. In contrast, when the field is along the c-axis, the
effect of the field is stronger and the divergence of the specific
heat is more easily suppressed as the field is increased. For a
comparable relative change in field of 41$\%$, corresponding to H
= 7~T in this orientation, $\gamma$(0.2~K) is readily suppressed
to 0.9~J/mol K$^{2}$ which is 25$\%$ less than its value at
H$_{c2}$ for the same temperature of 0.2~K. With further
increasing magnetic field in the c-axis orientation, the specific
heat tends to saturate at low temperatures and a clear Fermi
Liquid regime extends up to 0.5~K at the highest field of 9~T as
was reported
in ref.~\cite{AndreaQCP:prl-03}.\\

\begin{figure}
\includegraphics[width=3.3in]{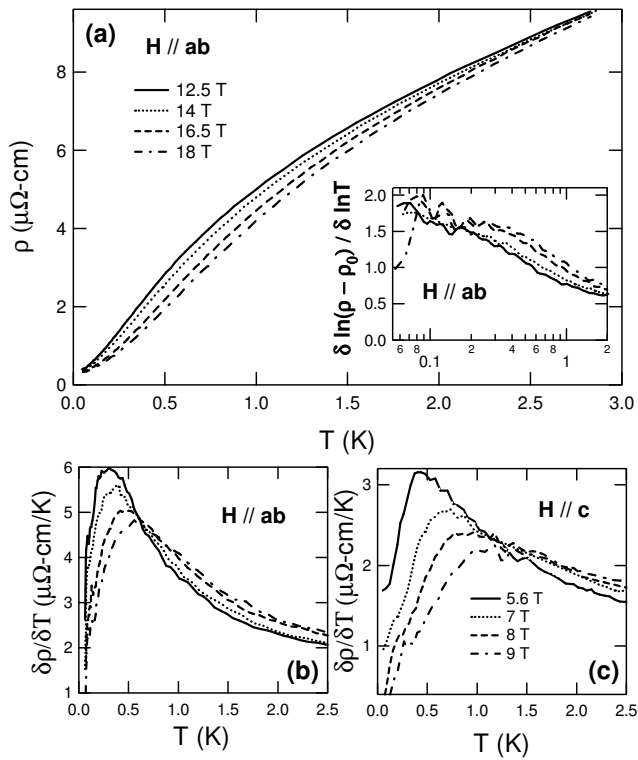}
\caption{\label{Resistivity-fig2} (a) Resistivity as a function of
temperature between 50~mK and 3~K for magnetic fields up to 18~T
oriented in the plane. Inset: the same data is shown as a semi-log
plot of the effective exponent of resistivity as a function of
temperature, defined as the logarithmic derivative of
$\rho-\rho_{0}$. Note that the asymptotic value at high
temperatures is less than 1. The maximum in $\frac{\delta
\rho}{\delta T}$ for H$\parallel$ab (panel b) and H$\parallel$c
(panel c) tracks the inflection point in resistivity, from which
one can see that the field has a significantly greater effect when
applied along the c-axis.}
\end{figure}

Resistivity as a function of temperature in a separate single
crystal measured between 50~mK and 3~K with magnetic fields
applied in-plane, from 12.5~T to 18~T is shown in
figure~\ref{Resistivity-fig2}(a). Electric current was applied
in-plane but perpendicular to the field. The overall S-shape of
the resistivity seen on the upper panel in this orientation is
qualitatively similar to the c-axis data published
previously~\cite{JP1:prl-03,AndreaQCP:prl-03}. In the low
temperature limit the curvature of resistivity is upward and the
negative magnetoresistance is significant, in contrast to the high
temperature regime where the curvature becomes negative and the
magnetoresistance is reduced. Despite the qualitative similarities
in the overall shape of $\rho$ for H$\parallel$c and
H$\parallel$ab, there is a striking difference in the rate of
evolution as a function of field between the two orientations.
This can be characterized by the temperature T$^{\ast}$ of the
inflection point in resistivity versus temperature, which will
appear as a maximum in $\frac{\delta \rho}{\delta T}$. Obviously
these temperatures are larger than the temperatures up to which
the Fermi Liquid behavior extends. Nevertheless, T$^{\ast}$ can be
taken as a crossover between the low temperature Fermi Liquid
regime and high temperature Non Fermi Liquid regime, as in the
case of YbRh$_{2}$Si$_{2}$~\cite{Custers:nature-03}.
Figures~\ref{Resistivity-fig2}(b) and~\ref{Resistivity-fig2}(c)
show that the temperature of the inflection point rises much more
rapidly with increasing field for H$\parallel$c than for
H$\parallel$ab.\\

The inset of figure~\ref{Resistivity-fig2}(a) presents the
logarithmic derivative of $\Delta \rho= \rho-\rho_{0}$ with
respect to temperature, as a function of temperature. The
evolution of this effective exponent is worth a few comments.
First, in a temperature range which is dominated by quantum
critical fluctuations as evidenced by the observed scaling in
refs.~\cite{AndreaQCP:prl-03} and~\cite{JP2:cond-mat-04}, the
exponent saturates to a value less than 1 in the high temperature
limit, independent of the field. This is consistent with the 2/3
exponent in this temperature range reported for field along the
c-axis at finite fields~\cite{JP2:cond-mat-04}. Second, the value
of 2 corresponding to Fermi Liquid regime is only reached in the
limit of low temperatures, with an onset temperature increasing
slightly with field. In the intermediate temperature range, one
observes a plateau around the value of 3/2 which becomes more
extended in temperature as the field is increased. The overall
shape of the logarithmic derivative is reminiscent of the
theoretical curves from ref.~\cite{Rosch:prb-00} in the framework
of a spin density wave scenario, and, in general, emphasizes that
there is no universal, single power law temperature dependence in
the Non Fermi Liquid behavior of resistivity.
\\

\begin{figure}
\includegraphics[width=3.3in]{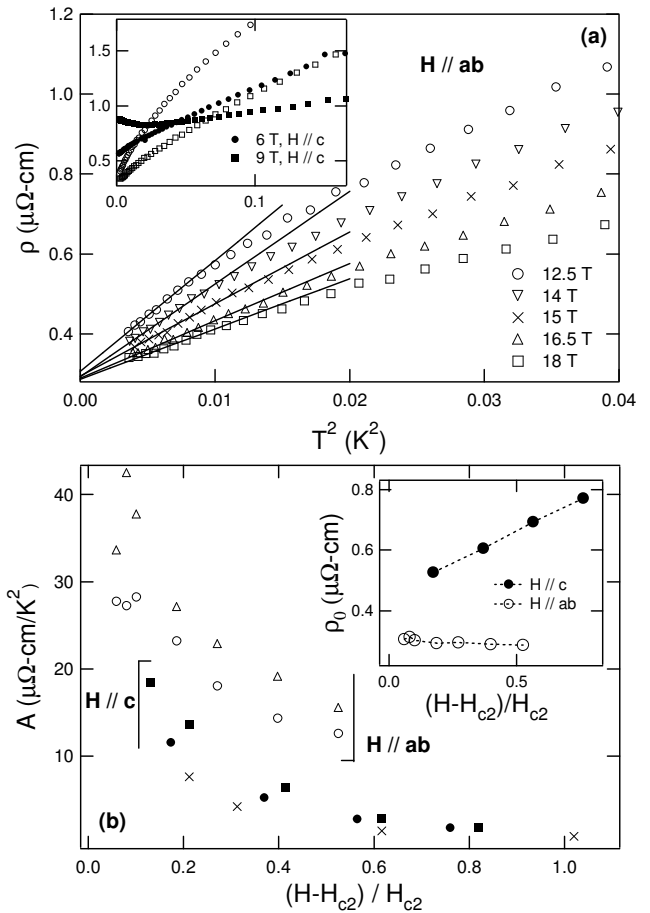}
\caption{\label{AvsH-fig3} (a) Resistivity vs. T$^{2}$ at low
temperatures for H$\parallel$ab. The symbols represent data
between 12.5~T and 18~T as indicated in the figure and the solid
lines represent the T$^{2}$ fit. The inset shows the 12.5~T and
18~T data for field in-plane (open symbols) together with 6~T and
9~T data for field along the c-axis (filled symbols) in the same
single crystal for comparison. Note the larger slope and smaller
temperature range for the T$^{2}$ behavior in resistivity when the
field is in-plane. (b) Coefficient $A$ as a function of reduced
field $\frac{H-H_{c2}}{H_{c2}}$ obtained from the
$\rho=\rho_{0}+AT^{2}$ fits for field parallel (filled symbols)
and perpendicular (open symbols) to the c-axis. The open ($\circ$)
and closed ($\bullet$) circles correspond to the data on the same
sample. The inset of the lower panel shows the residual
resistivity $\rho_{0}$ vs. reduced field for both orientations for
the same sample. In the main panel: $\triangle$ corresponds to a
second sample measured with field in-plane and parallel to the
electric current, $\blacksquare$ is for the sample of
Ref.~\cite{AndreaQCP:prl-03} and $\times$ is taken from
Ref.~\cite{JP1:prl-03} both with field parallel to the c-axis. All
data have current in the plane. Note that the A coefficient for
field in-plane is larger than the A coefficient for field along
the c-axis.}
\end{figure}

\begin{figure}
\includegraphics[width=3.3in]{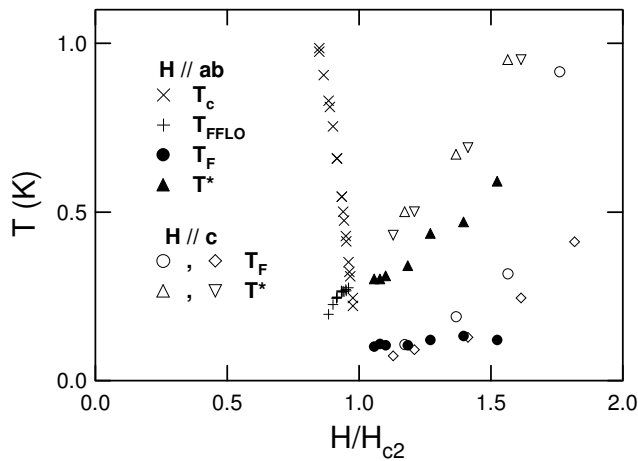}
\caption{\label{PhaseDiagram-fig5} H-T phase diagram of \Co~on a
reduced scale for both H$\parallel$c and H$\parallel$ab
orientations. The superconducting upper critical field is obtained
from previous specific heat measurements of
ref.~\cite{Andrea:prl-02,AndreaFFLO:prl-03}, T$_{F}$ was
determined from the T$^{2}$ resistivity fits, and T$^{*}$ is the
inflection point in resistivity versus temperature. T$_{F}$ and
T$^{*}$ for H$\parallel$ab are from the same sample whose data for
H$\parallel$c are shown with $\circ$ and $\triangle$ respectively.
Symbols $\diamond$ and $\triangledown$ are obtained from data
presented in ref.~\cite{AndreaQCP:prl-03}}
\end{figure}

The fact that the in-plane resistivity can be fitted with a
quadratic $\rho=\rho_{0}+AT^{2}$ law only at the lowest
temperatures is shown in the upper panel of
figure~\ref{AvsH-fig3}, corresponding to a Fermi Liquid regime
over a rather limited temperature range. This is consistent with
Non Fermi Liquid behavior in the specific heat data extending over
a large temperature range for the in-plane orientation, as
described above. As the magnetic field increases from 12.5~T to
18~T, the A coefficient corresponding to the slope of the T$^{2}$
behavior is significantly reduced. At the same time, the
temperature up to which the T$^{2}$ fit holds, defining the Fermi
temperature obtained from resistivity, increases slightly but
systematically~\cite{TFdefinition}. The magnetic field dependence
of the A coefficient and the Fermi temperature are displayed in
figure~\ref{AvsH-fig3}(b) and ~\ref{PhaseDiagram-fig5},
respectively. Both the enhancement of the A coefficient and the
decrease in the Fermi temperature are consistent with the specific
heat diverging to lower temperatures as the field approaches the
superconducting upper critical field. This suggests the presence
of a field-tuned quantum critical point in the vicinity of
H$_{c2}$ for field in-plane orientation. A similar conclusion has
been drawn in the previous reports for field along the
c-axis~\cite{JP1:prl-03,AndreaQCP:prl-03}. Thus, our results imply
that the quantum critical point has the same anisotropy as the
superconducting upper critical field.
\\

Despite the analogy between the two field orientations, one
notices a quantitative difference when comparing the rate at which
the magnetic field tunes the system into a Non Fermi liquid
regime. Not only is the Fermi Liquid regime restricted to a
smaller portion of the phase diagram but the rise of the Fermi
temperature as the field increases above H$_{c2}$ is more gradual
for the in-plane orientation, as shown in
figure~\ref{PhaseDiagram-fig5}. This is also directly seen in the
difference between resistivity for field in-plane (at 12.5~T and
18~T) and for field along the c-axis (at 6~T and 9~T) again in the
same crystal, as shown in the inset of figure~\ref{AvsH-fig3} (a).
Clearly, the quadratic temperature dependence of resistivity has a
stronger slope, when the field is in-plane, but the data deviates
from T$^{2}$ law at a much  lower temperature than the c-axis
data. The difference in the A coefficient for the two field
orientations is shown in figure~\ref{AvsH-fig3} (b). We have
compiled the results of the T$^{2}$ fits from various samples for
comparison, and present the A coefficient as a function of the
reduced field $\frac{H-H_{c2}}{H_{c2}}$ for the two orientations
(with H$_{c2}^{\parallel ab} = 11.8$~T and H$_{c2}^{\parallel c} =
4.95$~T). Included in figure~\ref{AvsH-fig3} are results from data
in figure~\ref{Resistivity-fig2}, with the magnetic field applied
parallel and perpendicular to the c-axis in the same crystal, as
are results from previously published data for H$\parallel$c from
ref.~\cite{JP1:prl-03,AndreaQCP:prl-03}. Moreover, we have
included results from longitudinal magnetoresistance data, with
field parallel to the current and parallel to the plane (raw data
not shown). We find that the field in-plane A coefficient is
systematically larger than the c-axis one, but it is less
divergent as well, beyond sample-to-sample variations. The
residual resistivity also has a different evolution depending on
the field orientation. The inset of figure~\ref{AvsH-fig3} (b)
shows that the residual resistivity is almost constant as a
function of the reduced field, when field is applied in the plane,
but it is strongly increasing when field is along the
c-axis in the same single crystal.\\

At this point, a word of caution regarding the analysis of the
resistivity data is in order. It is by no means clear that the
data down to 50~mK has saturated to its limiting T$^{2}$ behavior.
This becomes even more likely as the field approaches H$_{c2}$,
and the already limited range for T$^{2}$ behavior systematically
shrinks. An additional problem for the determination of the A
coefficient for H$\parallel$c can be due to the low temperature
upturn in resistivity. Even though we exclude this portion of the
data from the fits, the values are still underestimated when
compared to the sample of ref.~\cite{AndreaQCP:prl-03} where no
upturn was present. A similar upturn has also been reported in
ref.~\cite{JP1:prl-03} and was presumed to be a Fermi surface
effect involving closed orbits when the quantum limit is reached.
This is consistent with the two dimensional nature of parts of the
Fermi surface, as we do not observe the upturn in the same sample
when the magnetic field is in-plane, twice as large, and still
perpendicular to the current, rather only for H$\parallel$c, as
shown in the inset of figure~\ref{AvsH-fig3} (a). The trend in the
c-axis data is that the upturn becomes more pronounced and starts
at a higher temperature as the field increases, and the values are
consistent with the $\omega_{c}\tau=1$
condition~\cite{wctau_footnote}. So it is quite possible that the
reported A values may be lower bounds, which is why we have
refrained from quantitatively fitting the divergences. However, it
is clear that potential corrections, were we able to measure to
lower temperatures, would only increase the divergence of A, thus
making the case for the field tuned-quantum critical point to lie
at the
superconducting H$_{c2}$ even stronger.\\

We should also stress that the uncertainties related to the
analysis do not compromise the validity of the points we
emphasize. The inset of figure~\ref{AvsH-fig3} (a) shows that
there is indeed substantially more scattering when the field is
oriented in the plane. Our determination of the A coefficient,
which suggests a quantum critical point at the superconducting
H$_{c2}$, is corroborated by a log(T) divergence in C/T at
H$_{c2}$ down to the lowest temperatures measured by specific
heat. Furthermore, the fact that the non-Fermi liquid regime is
more robust when the field is applied in the plane is seen clearly
both in the specific heat data of
figure~\ref{Anisotropy-gamma-fig1} and in tracking the inflection
point in the resistivity curves in figure~\ref{Resistivity-fig2}.\\

In trying to understand our data, it is constructive to first
consider the effect of the Fermi surface topology. de Haas-van
Alphen (dHvA) measurements reveal that the Fermi surface of
\Co~has large 2-dimensional surfaces as well as small
3-dimensional pockets~\cite{Hall:prb-01,Settai:JPCM-01}. If the
conductivity is dominated by the 2-dimensional surfaces one would
expect an orbital component to the magnetoresistance which is
always positive when the field is along the c-axis (closed
orbits), and field independent with H$\parallel$ab (open
orbits)~\cite{Hurd:book}. Indeed, this explains why the residual
resistivity has a positive field dependence for H$\parallel$c, and
none for H$\parallel$ab. The field-independent in-plane residual
resistivity further suggests that the elastic scattering from
disorder is not affected by the strong fluctuations leading to the
large mass enhancement. At finite temperatures the negative
magnetoresistance is accounted for by the suppression of quantum
fluctuations (and hence the A coefficient) as one moves away from
the quantum critical point. This is also seen in dHvA by the
reduction of the effective mass of the 2-dimensional sheets as the
field is increased beyond H$_{c2}$ with
H$\parallel$c~\cite{Settai:JPCM-01}. The fact that the masses of
the 3-dimensional pockets are significantly less enhanced as the
critical field is approached suggests that transport with the
current along the c-axis should be markedly
different~\cite{Artur:unpublished}. Unfortunately, dHvA is blind
to the 2 dimensional sheets when the field is oriented in the
plane, and so it can not compare the relative mass enhancement for
the two field orientations as we have done here. Moreover, the
anisotropy of the spin fluctuations with respect to the field
orientation is yet to be established by a direct probe like NMR or
neutron scattering.
\\

Can an anisotropic g-factor, which represents the effective
coupling between the magnetic excitations and the external field,
explain the observed anisotropy in the scattering rates? By
fitting the H$_{c2}$(T) curves Maki {\it et al.} found g values of
1.5 and 0.64 for H$\parallel$c and H$\parallel$ab, respectively
which is nearly identical to the ratio of H$_{c2}$ for the two
field orientations~\cite{Maki:prb-04}. Thus, we have attempted to
take into account the anisotropy in the g-factor (and in H$_{c2}$)
by plotting the data against a renormalized H.  The phase diagram
as a function of temperature and reduced
field~$(\frac{H}{H_{c2}})$, shown in
figure~\ref{PhaseDiagram-fig5}, nearly accounts for the anisotropy
in the inflection point of the resistivity curves. However, it
does not account for the anisotropy in A(H) from
figure~\ref{AvsH-fig3} (b). Accounting for different g-factors,
close to a critical point one expects the A coefficient to diverge
as A(H) = A$_{0}$ $(\frac{H-H_{c2}}{H_{c2}})^{\alpha}$.
Figure~\ref{AvsH-fig3} (b) would then lead us to conclude that
A$_{0}^{H\parallel ab}$ $\approx$ 3A$_{0}^{H\parallel c}$. In
addition, the Fermi temperature anisotropy is also not accounted
for in this way. While the values for H$\parallel$ab may simply
represent upper limits, the values for H$\parallel$c of 8~T and
9~T are well established by both specific heat and resistivity.
Thus we can see that for $\frac{H}{H_{c2}} > 1.5$,
T$_{F}^{\parallel ab}$ is less than half T$_{F}^{\parallel c}$.
This is also consistent with our discussion of the anisotropy in
the specific heat data of figure~\ref{Anisotropy-gamma-fig1}.
Thus, we conclude that the quantum fluctuations are significantly
more robust when the field is applied in the plane, and that the
origin of this anisotropy is not solely a result of an anisotropic
g-factor.
\\

At this point, we consider how our data impact the various quantum
critical point scenarios for \Co. One possibility is that the
quantum critical behavior originates from the second
superconducting phase which was identified for
H$\parallel$ab~\cite{AndreaFFLO:prl-03}. Although the inflection
point in the resistivity curves appear to originate from this
phase boundary for H$\parallel$ab, there is no other data which
ties this phase transition to the quantum critical behavior. In
addition, for H$\parallel$c the inflection point in resistivity
has nearly identical behavior to H$\parallel$ab, while a second
superconducting transition is strongly suppressed for this
orientation~\cite{AndreaFFLO:prl-03}. Thus we can confidently rule
out this origin for the quantum
critical behavior.\\

Why is the quantum critical field tied so closely to the
superconducting H$_{c2}$? We now believe this to be more than a
mere coincidence since attempts to separate one from the other
with either Sn doping~\cite{Eric:unpublished} or field orientation
(more than a factor of 4 change in H$_{c2}$ combined) could not do
so. Thus, it would appear that the quantum critical behavior
originates from a superconductor to paramagnet quantum phase
transition. However, the width of the fluctuation regime for a BCS
superconductor is extremely small, even for nodal superconductors,
and can be approximated by the Ginzburg criteria to be $\Delta
T/T_{c}~= [(2\pi\xi_{0})^{-3}k_{B}/\Delta C]^{2}~\approx 10^{-9}$
for \Co. Disorder can increase this fluctuation regime by pair
breaking effects possibly leading to a quantum critical point as
shown in ref.~\cite{Ramazashvili:prl-97}, but we would not expect
this to apply to the extremely pure system of \Co. Further, we
note that the superconducting H$_{c2}$(T) boundary as
T~$\rightarrow$~0 is first order~\cite{Andrea:prl-02}. Thus it
would require a truly novel type of superconductor to produce the
observed quantum critical point. An alternative view is that the
quantum criticality in this system originates from an
antiferromagnetic quantum critical point. Then the correct
question to ask is why is the superconducting H$_{c2}$ tied to the
quantum critical point? This could be possible if the low field
phase had a large susceptibility to become superconducting.
Superconductivity is then destroyed at the quantum phase
transition since the susceptibility to superconductivity in the
high-field phase is significantly lower. This is precisely what
has been observed theoretically in low density
systems~\cite{Howell:cond-mat-01}. In principle the quantum
critical point could separate any two ground states, but comparing
\Co~to \Rh~suggests that the quantum critical point separates an
antiferromagnetic ground state from a high field paramagnetic
state. For the case of an antiferromagnetic quantum critical point
we might also expect to find short range antiferromagnetic order
inside the vortex cores below H$_{c2}$.
\\

In conclusion, we have measured specific heat and resistivity in
\Co~with H$\parallel$ab. The specific heat shows C/T $\propto$
log(T) down to the lowest temperature measured at H$_{c2} \simeq
12$~T. Resistivity measurements also show that the
electron-electron scattering diverges at H$_{c2}$, and that at
high temperatures the resistivity has a sub-linear power law.
Thus, for both field orientation there is a field-tuned quantum
critical point close to H$_{c2}$, but the Fermi temperature is
smaller and the tuning much slower for the field in-plane
orientation. This means that the magnetic field is more efficient
in suppressing the heavy fermion ground state in the c-axis
orientation. The fact that it is experimentally impossible to
distinguish the quantum critical point from the upper critical
field independent of field orientation must be a consequence of a
common underlying mechanism for both phenomena. The origin of the
anisotropy in the tuning rate with respect to the field
orientation might provide a clue to
the nature of the fluctuations that become critical at H$_{c2}$.\\

\begin{acknowledgments} We acknowledge fruitful discussions with A.
Abanov, E.D. Bauer, N. Curro, A. Millis, A. Rosch, A. Schofield,
and I. Vekhter. F.R. thanks the Reines Postdoctoral Fellowship
(DOE-LANL) for support. Work at Los Alamos National Laboratory was
performed under the auspices of the U.S. Department of Energy.
Work at the NHMFL was performed under the auspices of the National
Science Foundation, the State of Florida and the US Department of
Energy.
\end{acknowledgments}


\begin{thebibliography}{25}
\bibitem[{\citenamefont{TFdefinition}(2004)}]{TFdefinition}
\bibinfo{note}{T$_{F}$ was defined as the temperature at which there
was a 2$\%$ deviation of the data from the T$^{2}$ fit}.
\bibitem[{\citenamefont{wctau_footnote}(2004)}]{wctau_footnote}
\bibinfo{note}{If we let $\omega_{c}= \frac{eH}{m^{*}c}$, $\tau=
\frac{m^{*}}{ne^{2}\rho}$, and $n=6.2\times10^{27}~m^{-3}$ for
CeCoIn$_{5}$ then $\omega_{c}\tau=1$ implies $\rho[\mu \Omega
.cm]=0.1010199 \times H$[T]. For 7~T and 9~T  this gives $\rho
\simeq 0.7 \mu \Omega .cm$ and $\rho \simeq 0.9 \mu \Omega .cm$
respectively, which is in good agreement with the experimental
value of the resistivity minimum in the H$\parallel$c
orientation}.
\end{thebibliography}
\end{document}